\documentclass[11pt]{article}

\usepackage[margin=1in]{geometry}
\usepackage{amsmath, amssymb, amsfonts}
\usepackage{graphicx}
\usepackage{placeins}
\usepackage{float}

\usepackage{microtype}
\usepackage{booktabs}
\usepackage{array}
\usepackage{caption}
\usepackage{xcolor}
\usepackage{hyperref}

\hypersetup{
  hidelinks,
  pdftitle={Hamiltonian-Dependent Compressibility and Physical Preservation in Approximate Quantum Compilation using Tensor Networks},
  pdfauthor={Dohyun Kim and Namshik Han}
}

\newcommand{\makepreprinttitle}{%
\begin{center}
\vspace*{-1.8em}
{\LARGE\bfseries Hamiltonian-Dependent Compressibility and Physical Preservation in Approximate Quantum Compilation using Tensor Networks\par}
\vspace{1.0em}
{\large Dohyun Kim\(^{1}\) and Namshik Han\(^{1,2,3,4,5,6,7,8,9,*}\)\par}
\vspace{0.45em}
{\small
\begin{minipage}{0.98\textwidth}
\centering
\(^{1}\)Department of Artificial Intelligence, Yonsei University, Seoul, Republic of Korea\par
\(^{2}\)Department of Quantum Information, Yonsei University, Seoul, Republic of Korea\par
\(^{3}\)Institute for Convergence Research and Education in Advanced Technology, Yonsei University,\\
Seoul, Republic of Korea\par
\(^{4}\)Department of Nano Biomedical Engineering (NanoBME), Advanced Science Institute, Yonsei University,\\
Seoul, Republic of Korea\par
\(^{5}\)Center for Nanomedicine, Institute for Basic Science (IBS), Seoul, Republic of Korea\par
\(^{6}\)College of Pharmacy, Yonsei University, Seoul, Republic of Korea\par
\(^{7}\)Milner Therapeutics Institute, University of Cambridge, Cambridge, United Kingdom\par
\(^{8}\)Cambridge Centre for AI in Medicine, University of Cambridge, Cambridge, United Kingdom\par
\(^{9}\)Cambridge Stem Cell Institute, University of Cambridge, Cambridge, United Kingdom\par
\end{minipage}\par
}
\vspace{0.4em}
{\footnotesize Email: \href{mailto:ts2526@yonsei.ac.kr}{ts2526@yonsei.ac.kr}; \(^{*}\)Corresponding author: \href{mailto:namshik.han@yonsei.ac.kr}{namshik.han@yonsei.ac.kr}\par}
\end{center}
\vspace{-0.4em}
}
\begin{document}

\makepreprinttitle

\begin{abstract}
Approximate quantum compilation (AQC) using tensor networks provides a promising route to reducing the circuit resources required for quantum simulation. However, the factors governing the compressibility of many-body dynamics and the extent to which compressed circuits retain physically relevant information on quantum hardware remain poorly understood. We investigate these questions using AQCtensor for an 80-qubit one-dimensional anisotropic XYZ Heisenberg chain initialized in a N\'{e}el state. By varying the exchange anisotropy and external magnetic field, we find that compression performance changes systematically with the anisotropic coupling configuration and field orientation. These variations are strongly associated with a nested-commutator measure related to second-order Trotter error, with corresponding directional trends also reflected in the target-state MPS bond dimension. For an 80-qubit quantum-hardware evaluation, AQCtensor compresses a 10-repetition target evolution to 5 repetitions while retaining a target-state fidelity of 0.995474, reducing the total circuit depth by 45.9\% and the two-qubit gate count by 47.6\%. The compressed circuit exhibits comparable behavior across magnetization, nearest-neighbor correlation, and Hamiltonian expectation measurements, while process-level diagnostics and local reduced-state tomography provide complementary characterization beyond these observables. Together, these results link AQCtensor compressibility to Hamiltonian commutator structure and target-state complexity and demonstrate substantial circuit reduction while retaining physically relevant features in an 80-qubit quantum-hardware experiment.
\end{abstract}


\vspace{1em}

\section{Introduction}
\label{sec:intro}

Quantum simulation is one of the most promising applications of quantum computing, offering a direct framework for studying the dynamics of interacting many-body systems \cite{RevModPhys.86.153}. On gate-based qubit processors, however, continuous time evolution under a Hamiltonian $H$, 

\[
U(t)=\exp(-iHt)
\]

must be represented by a finite sequence of quantum gates. Product-formula methods, including Lie--Trotter and Suzuki--Trotter decompositions, provide a widely used approach by approximating the evolution as repeated short-time propagations \cite{Childs2019NearlyOL}.
Increasing the number of product-formula steps generally improves the approximation to the target dynamics, but at the cost of greater circuit depth and a larger number of two-qubit gates. On present-day quantum hardware, this trade-off is particularly restrictive because deeper circuits accumulate more hardware errors and can become impractical before the desired simulation accuracy is reached.

Approximate quantum compilation (AQC) offers a complementary strategy by directly replacing a deep target circuit with a shallower variational circuit optimized to reproduce its output state \cite{Khatri2018QuantumassistedQC,Madden2022Best}. Direct optimization of global cost functions, such as state-overlap objectives, can nevertheless encounter barren plateaus as the system size increases \cite{Cerezo2021CostFD,Robertson2022Escaping}. 
Tensor-network-based AQC methods are particularly attractive for many-body quantum simulation because classically tractable matrix product state (MPS) representations can be used during the optimization. AQCtensor implements this idea by representing the target and variational states as MPSs and optimizing a lower-depth circuit against the target evolution \cite{aqc-tensor-research}. Previous works have shown that such approaches can substantially reduce circuit depth while maintaining high state fidelity and can be extended to large many-body systems and quantum-hardware demonstrations \cite{Jaderberg2026Variational,Pennington2026Preparing}.

These results establish that substantial compression is possible, but why some many-body evolutions are more compressible than others remains less clear. A natural candidate is the complexity of the target
state itself. Because AQCtensor represents the target state as an MPS, the MPS
bond dimension provides a convenient measure of this complexity
\cite{Verstraete2006MatrixPS}. However, target-state complexity alone may not
fully determine compressibility: states with similar MPS bond dimensions can
still differ in how easily they are reproduced by a shallow Trotter-based
ansatz. This suggests that additional structure, beyond the target state
itself, may influence the difficulty of compression.

Another possibility is that compressibility depends on the relationship
between the Hamiltonian structure and the shallow product-formula ansatz.
Product-formula errors are governed by the noncommutativity of Hamiltonian
terms through commutators and nested commutators
\cite{PhysRevX.11.011020,Casas2025Commutators}. Because AQCtensor is
initialized from a shallow product-formula circuit, the same commutator
structure may influence how effectively the variational ansatz can reproduce
the deeper target evolution. This suggests that compressibility may depend
not only on target-state complexity, but also on the compatibility between
the Hamiltonian dynamics and the shallow ansatz. This connection has not yet
been systematically examined.

Beyond the factors governing compressibility, the extent to which compression preserves the physically relevant features of the original many-body evolution on quantum hardware remains an open question. Previous AQC-based hardware studies have demonstrated physically meaningful observables and many-body signatures \cite{Jaderberg2026Variational,Pennington2026Preparing}, but preservation across observable, state, and process levels remains less systematically characterized. 

In this work, we investigate both the factors associated with AQCtensor
compressibility and the physical information retained after compression in an
80-qubit one-dimensional anisotropic XYZ Heisenberg chain. We first examine
how compression performance varies across anisotropic coupling and
external-field configurations, and relate these variations to Hamiltonian
commutator structure and target-state MPS complexity. We then test whether these compressed dynamics remain physically meaningful
at scale by executing the original and compressed circuits on 80 qubits of
quantum hardware, using physical observables, local reduced-state tomography,
and process-level diagnostics.

The remainder of this paper is organized as follows. Sec.~\ref{sec:framework} describes the AQCtensor compression procedure, Hamiltonian and simulation settings, and hardware-evaluation methods. Sec.~\ref{sec:compress} presents the Hamiltonian-dependent compressibility results and analyzes their relation to commutator structure and MPS complexity. Sec.~\ref{sec:hardware} quantifies the circuit-resource reduction achieved by AQCtensor and evaluates physical preservation through hardware observables, local-state analysis, and process-level diagnostics. Sec.~\ref{sec:discussion} discusses the implications and limitations of these results.

\section{Methods}
\label{sec:framework}

\subsection{AQCtensor compression}

Circuit compression was performed using the AQCtensor workflow implemented in qiskit-addon-aqc-tensor \cite{qiskit-addon-aqc-tensor}, following the standard procedure described in the original AQCtensor work \cite{aqc-tensor-research}. The target and variational states were represented as matrix product states (MPSs), and shallower Trotter circuits were optimized against the target evolution. The ansatz parameters were optimized by minimizing the state infidelity using L-BFGS-B with gradients computed by automatic differentiation in JAX. Optimization was terminated when the L-BFGS-B convergence criterion was satisfied or when the maximum of 200 iterations was reached. Among the optimized candidates, we selected the shallowest circuit achieving a target-state fidelity of at least $F \ge 0.99$.

\subsection{Hamiltonian and simulation settings}

We consider an open 80-qubit one-dimensional anisotropic XYZ Heisenberg chain initialized in the N\'{e}el state
$|\phi_0\rangle=|0101\cdots\rangle$.
We use a second-order Suzuki--Trotter product formula for the main compression experiment and the Hamiltonian-parameter sweep, with different simulation settings for each.

For the main 80-qubit compression experiment, we use
$\mathbf{J}=(0.5\pi,1.0\pi,1.5\pi)$
and
$\mathbf{h}=(0.5\pi,1.0\pi,2.0\pi)$,
with a 10-repetition target circuit at total evolution time $t=0.30$.
This configuration was selected to produce a sufficiently pronounced observable signal, allowing differences between the original and compressed evolutions to be resolved more clearly on quantum hardware.

For the compressibility analysis, we perform a systematic Hamiltonian sweep to examine how compression difficulty changes with the underlying dynamics. We consider
$\mathbf{J}_1=(0.5 \pi,1.0 \pi,1.5 \pi)$
and
$\mathbf{J}_2=(1.0 \pi,0.5 \pi,1.5 \pi)$,
which interchange $J_x$ and $J_y$ while keeping $J_z$ fixed, and vary the strength and orientation of the external field. Each target evolution in this sweep is generated using a 20-repetition circuit at the same total evolution time $t=0.30$. Detailed field configurations are provided in Supplementary Method A.

For the auxiliary relative-circuit diagnostics in Fig.~\ref{fig:process}, we use $\mathbf{J}=(\pi/8,\pi/4,\pi/2)$ and $\mathbf{h}=(\pi/3,\pi/6,\pi/9)$, with first-order Lie--Trotter $2\to1$ compression at $t=0.02$. This shallower setting was chosen to reduce the hardware noise limitations encountered with deeper relative circuits.

\subsection{Hardware evaluation}
\label{sec:protocol}

The original and AQCtensor-compressed circuits were evaluated on IBM Quantum's \texttt{ibm\_boston} backend, based on a Heron r3 processor, using observable-level measurements, local reduced-state tomography, and relative-circuit diagnostics.

For the observable-level hardware benchmarks, we evaluated both the original and compressed circuits together with self-mirrored versions constructed by appending the inverse of each circuit. For a circuit $U$, the resulting sequence $U^\dagger U$ ideally reduces to the identity and provides a complementary benchmark of accumulated hardware error \cite{Proctor2021ScalableMirror}. The Néel-state preparation is included in $U$ throughout this construction.

We measured the mean longitudinal and transverse magnetizations,
\[
\langle Z\rangle=\frac{1}{N}\sum_{i=1}^{N}\langle Z_i\rangle,
\qquad
\langle X\rangle=\frac{1}{N}\sum_{i=1}^{N}\langle X_i\rangle,
\]
the mean nearest-neighbor correlation,
\[
C_{ZZ}=\frac{1}{N-1}\sum_{i=1}^{N-1}\langle Z_i Z_{i+1}\rangle,
\]
and the Hamiltonian expectation value $\langle H\rangle$ for the original and compressed circuits and their mirrored counterparts. Noiseless observable estimates  were obtained using Pauli propagation in the Heisenberg picture with the Qiskit pauli-prop package \cite{pauli-prop}.

Hardware measurements were further evaluated using error-mitigation techniques including readout mitigation, dynamical decoupling, twirling, and probabilistic error amplification with zero-noise extrapolation \cite{Cai2023QuantumEM,PhysRevA.105.032620,Wallman2016NoiseTF,Kim2023EvidenceFT}. Detailed quantum hardware evaluation and error mitigation procedures are provided in Supplementary Method B.

To characterize preservation beyond individual observables, we used process- and
local-state-level diagnostics. We constructed the relative circuit
$W = U_{\mathrm{original}}^\dagger U_{\mathrm{AQC}}$,
which approaches the identity as the two ideal evolutions become equivalent. Because direct characterization of the 80-qubit relative circuit is not tractable at this scale, mirror-circuit fidelity estimation (MCFE) was used to estimate the process fidelity of its hardware implementation \cite{Proctor2021ScalableMirror,Proctor2022EstablishingTrust}. We further evaluated the mean $Z$ expectation value of the relative circuit and reconstructed representative four-qubit reduced density matrices, which were compared using reduced-state fidelity and purity \cite{Xin2017RDM,Guo2023QuantumST}. Detailed procedures are provided in Supplementary Method C.

\section{Hamiltonian-dependent compressibility}
\label{sec:compress}

\subsection{Dependence on field strength and direction}

We first asked whether the degree of AQCtensor compression changes systematically with the underlying Hamiltonian. To test this, we varied the strength and direction of the external field for two anisotropic coupling configurations that differ only by exchanging $J_x$ and $J_y$.

\begin{figure}[htbp]
\centering
\includegraphics[width=.95\textwidth]{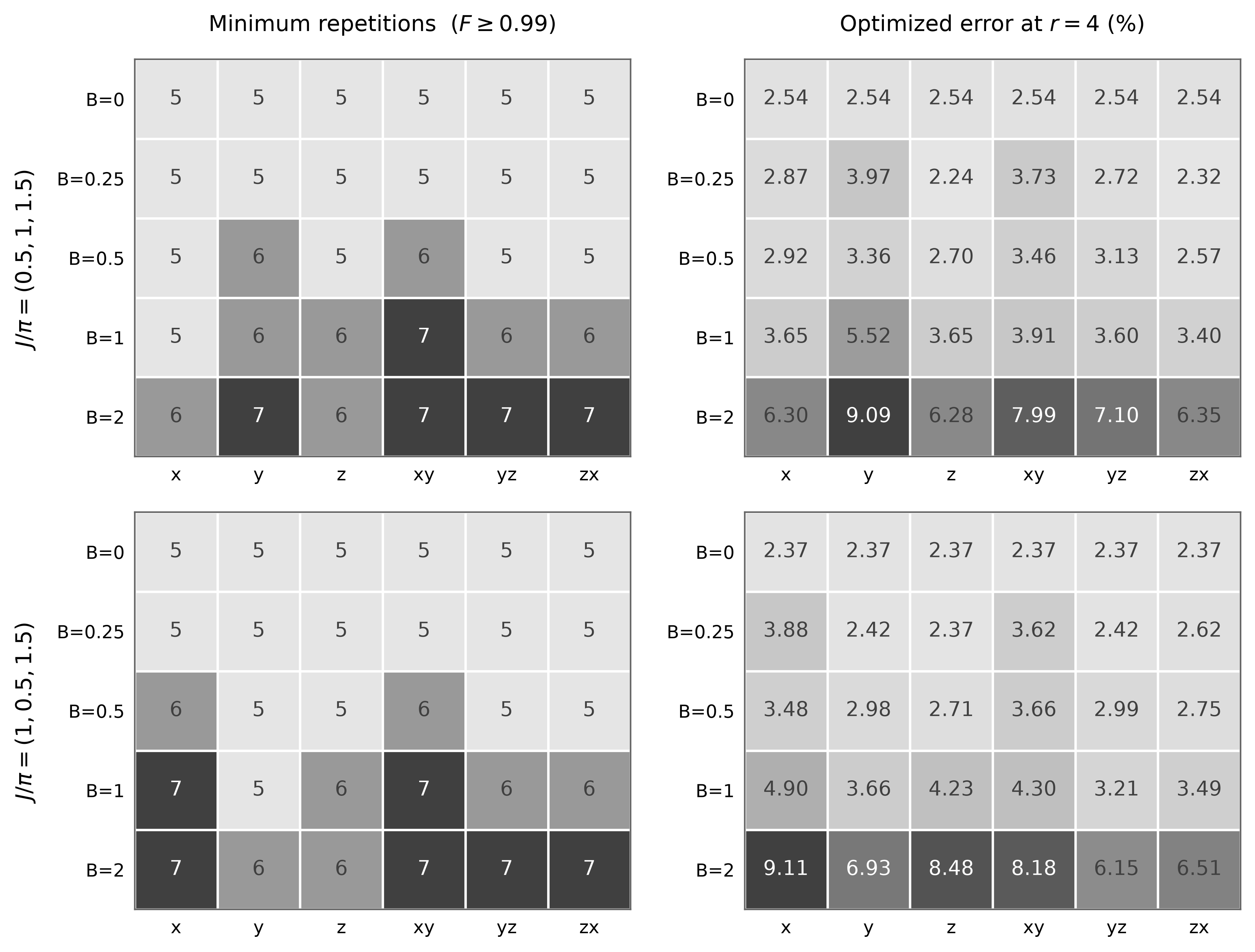}
\caption{AQCtensor compression performance across Hamiltonian parameters. Heatmaps show the minimum repetition count required to reach $F \geq 0.99$ and the optimized infidelity at fixed depth $r=4$ under the common
optimization protocol for different external field strengths and directions. At $B=0$, the $x$-, $y$-, and $z$-field entries correspond to the same zero-field Hamiltonian and therefore have identical values within each
coupling configuration.}
\label{fig:sweep}
\end{figure}

Fig.~\ref{fig:sweep} summarizes the resulting compression landscape. We quantify compression in two complementary ways: the minimum circuit repetition count required to reach $F \ge  0.99$, and the optimized infidelity $1-F$ at a fixed shallow depth of $r=4$. Lower repetition counts and lower infidelity therefore correspond to easier compression. Because the fixed-depth optimization was performed under a finite iteration
budget, the reported infidelities may be affected by incomplete convergence
and numerical variability in the optimization and should not be interpreted
as exact variational minima. We therefore do not rely on small quantitative
differences between symmetry-related configurations, but instead focus on
the broader variations observed across the Hamiltonian parameter space.

The compression performance varies with both field strength and field
direction. At weak fields, most Hamiltonian configurations exhibit similar
compression behavior, whereas stronger fields produce more pronounced
differences among field orientations.

For $J_1=(0.5\pi,1.0\pi,1.5\pi)$, the $y$-directed field shows the largest
fixed-depth infidelity among the single-axis cases at strong field.
At $B=2$, it requires $r=7$ to reach $F\geq0.99$ and yields a fixed-depth
infidelity of 9.09\% at $r=4$. After exchanging $J_x$ and $J_y$, the relative
$x$- and $y$-field responses are correspondingly interchanged, with the
$x$-directed field yielding an infidelity of 9.11\% at $B=2$.
This interchange is consistent with the symmetry relating the two coupling
configurations. Additional calculations with \(J_z/\pi=0.5\) show that the symmetry-related interchange of the \(x\)- and \(y\)-field responses persists, while the magnitude of the directional variation remains parameter dependent (Fig.~\ref{fig:supp_jz05}).

\subsection{Commutator structure and compressibility}
\label{sec:commutator}

The results in Fig.~\ref{fig:sweep}  show that AQCtensor performance varies with field
orientation across the anisotropic coupling configurations considered here.
In particular, exchanging $J_x$ and $J_y$ produces a corresponding
interchange in the relative $x$- and $y$-field responses, consistent with
the symmetry relating the two coupling sets. We therefore examine whether
these variations can be related to the commutator structure governing the
shallow Suzuki--Trotter representation.

The quality of the initial shallow Suzuki--Trotter circuit is relevant to the subsequent variational optimization. Previous AQCtensor studies have shown that convergence can depend strongly on initialization, with more accurate initial states providing more favorable starting points for optimization \cite{Jaderberg2026Variational}. We therefore ask whether the Hamiltonian commutator structure is associated with the difficulty of compressing the target evolution.

For second-order product formulas, the leading error is governed
by nested commutators of the Hamiltonian components rather than
by the first-order commutator alone \cite{PhysRevX.11.011020}.
More generally, higher-order corrections in product-formula
expansions are constructed from increasingly nested commutators
\cite{Casas2025Commutators}.
Motivated by this commutator-scaling theory, we introduce a
nested-commutator diagnostic to investigate the relationship
between Hamiltonian noncommutativity and AQCtensor compressibility.

For the implemented second-order Suzuki--Trotter decomposition,
we separate the Hamiltonian into the even-bond, odd-bond,
and field components,
\[
H = H_{\mathrm{even}} + H_{\mathrm{odd}} + H_{\mathrm{field}}.
\]
Here, $H_{\mathrm{even}}$ and $H_{\mathrm{odd}}$ contain the
$XX$, $YY$, and $ZZ$ interaction terms on alternating
nearest-neighbor bonds, respectively, while $H_{\mathrm{field}}$
contains the single-qubit external-field terms.

We then define the nested-commutator strength as
\[
C_2 =
\left[
\sum_{a,b,c\in\{\mathrm{even},\mathrm{odd},\mathrm{field}\}}
\left\|[H_a,[H_b,H_c]]\right\|_{2}^2
\right]^{1/2},
\]
where $\|O\|_2=(\sum_P |c_P|^2)^{1/2}$
denotes the Pauli-coefficient $L_2$ norm for
$O=\sum_P c_P P$.

This definition accounts for both bond--bond and
interaction--field noncommutativity, including contributions
that remain nonzero in the absence of an external field.
The resulting $C_2$ characterizes the aggregate
nested-commutator strength rather than the exact
Suzuki--Trotter error coefficient.
We compare this choice with the leading Baker-Campbell-Hausdorff (BCH) coefficient of the implemented
Suzuki--Trotter formula and an interaction--field-only commutator measure in Supplementary Method E.

Fig.~\ref{fig:mps}(a) and Fig.~\ref{fig:mps}(b) use the same Hamiltonian configurations as the compressibility sweep in Fig.~\ref{fig:sweep}. Fig.~\ref{fig:mps}(a) compares $C_2$ with the optimized AQCtensor infidelity at the fixed shallow depth $r=4$. Across the sampled Hamiltonian configurations, the two quantities show a strong positive association, with Pearson $r=0.887$ and Spearman $\rho=0.900$. Since both $C_2$ and the infidelity also vary with the field strength $B$, we additionally examined their correlation within fixed-$B$ subsets. The association remains positive within fixed-$B$
subsets, with Pearson $r=0.897$, $0.849$, $0.940$,
and $0.823$ and Spearman $\rho=0.478$, $0.478$,
$0.717$, and $0.717$ for $B=0.25$, $0.5$, $1$,
and $2$, respectively. These results indicate that the association between $C_2$ and the achieved fixed-depth AQCtensor infidelity remains positive when the field strength is
held fixed. The $B=0$ entries were excluded from the fixed-$B$ analysis because all field orientations correspond to the same zero-field Hamiltonian.
\label{sec:mps}

Fig.~\ref{fig:mps}(b) compares the same commutator measure with the target-state MPS complexity. For this analysis, the target-state MPS bond dimensions were evaluated using a truncation cutoff of $10^{-13}$. The association is positive but weaker than for
AQCtensor infidelity (Pearson $r=0.744$,
Spearman $\rho=0.454$). The weaker correlation indicates that, although Hamiltonian commutator structure is reflected in target-state MPS complexity, the latter alone does not fully account for the observed variation in AQCtensor compressibility.

To determine whether the directional trend extends beyond the two coupling configurations used in Fig.~\ref{fig:sweep}, we considered a broader family of anisotropic couplings by varying $J_0$, $J_z$, and the difference between $J_x$ and $J_y$. We then compared the differences in nested-commutator strength and target-state MPS bond dimension between the $x$- and $y$-field cases at $B=1$, $2$, and $3$.
 
For each Hamiltonian, we evaluated the directional differences

\[
\Delta C_2 = C_{2,y} - C_{2,x}
\]

and

\[
\Delta \chi_{\mathrm{MPS}} = \chi_y - \chi_x.
\]

As shown in Fig.~\ref{fig:mps}(c), the directional
changes in $C_2$ and MPS bond dimension become
increasingly aligned at larger field strengths,
with Pearson correlations of $r=0.537$, $0.925$,
and $0.959$ for $B=1$, $2$, and $3$,
respectively. In other words, when the nested-commutator structure shifts so that the $y$-directed evolution has a larger $C_2$ than the $x$-directed evolution, the MPS bond dimension tends to shift in the same direction, and vice versa. This directional agreement extends the observation in Fig.~\ref{fig:sweep} beyond the original two coupling configurations and shows that variations in the commutator structure are accompanied by corresponding variations in target-state complexity.

These results show that the Hamiltonian commutator structure is associated
with the observed variation in fixed-depth AQCtensor performance. Larger nested-commutator contributions are expected to produce larger finite-depth Suzuki–Trotter errors. This is relevant to AQCtensor because the shallow product-formula circuit provides the initialization for variational optimization.
Consistent with this interpretation, $C_2$ tracks the variation in the
achieved fixed-depth infidelity across the sampled Hamiltonian
configurations, while its directional variation is also reflected in the
target-state MPS bond dimension. These observations suggest that
commutator-based quantities capture a structural factor associated with
compressibility that is not described by state complexity alone.

\begin{figure}[htbp]
\centering
\includegraphics[width=.90\textwidth]{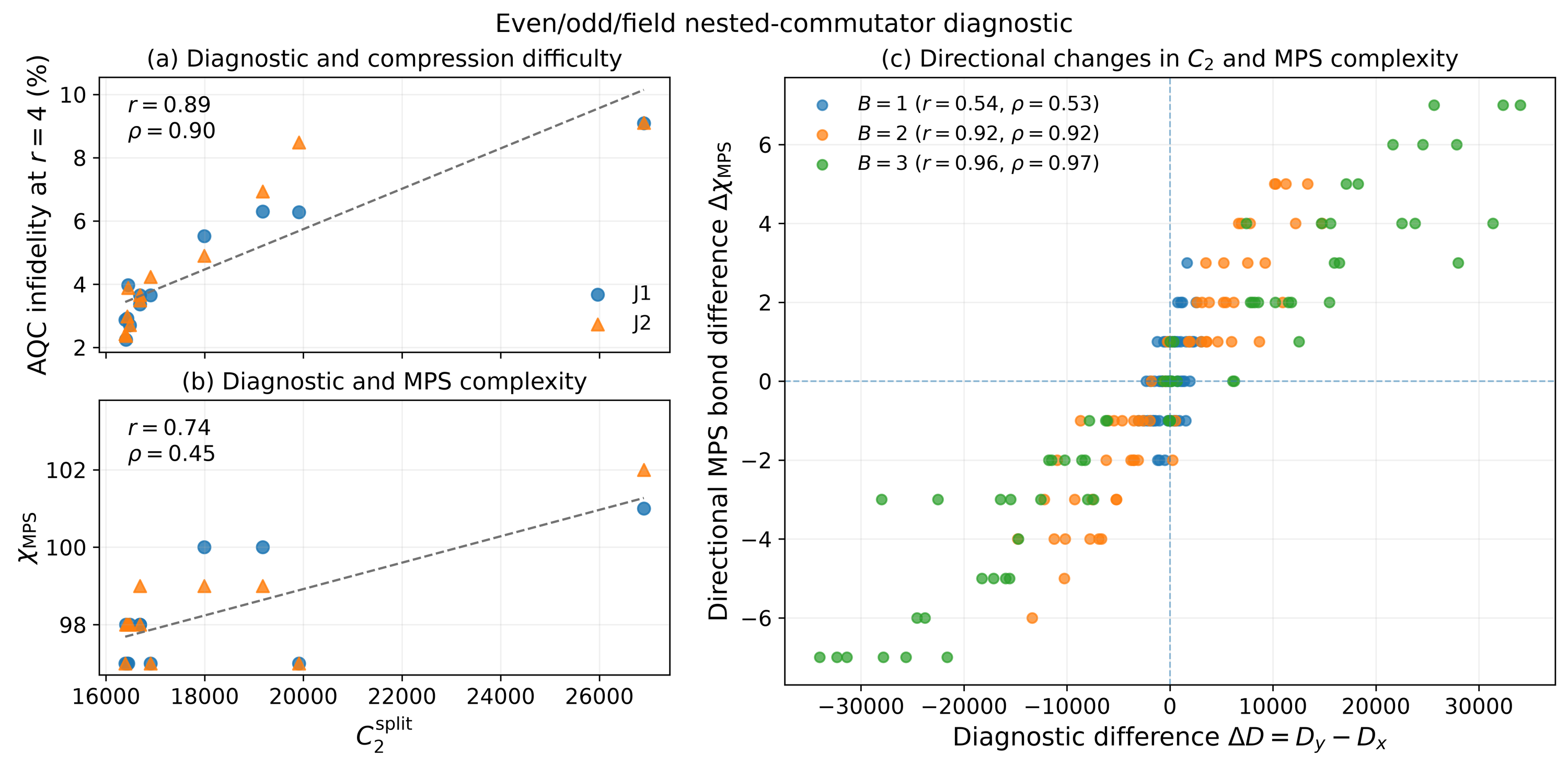}
\caption{Relation between Hamiltonian commutator structure, AQCtensor compression difficulty, and target-state MPS complexity. (a) AQCtensor infidelity at $r=4$ versus nested-commutator strength $C_2$. (b) MPS complexity versus $C_2$. (c) Directional changes in MPS complexity and $C_2$ across coupling and field configurations. }
\label{fig:mps}
\end{figure}

\section{Physical Preservation}
\label{sec:hardware}

\subsection{Circuit-resource reduction}
\label{sec:resource}

We next evaluated the practical effect of AQCtensor compression for the 80-qubit anisotropic XYZ Heisenberg chain considered in the hardware-validation setting. The original evolution is represented by a 10-repetition second-order Suzuki--Trotter circuit, while AQCtensor searches for a shallower circuit that reproduces the same target evolution with high fidelity. We first quantify the reductions in circuit depth and two-qubit gate count before examining the corresponding hardware results.

\begin{figure}[htbp]
\centering
\includegraphics[width=.95\textwidth]{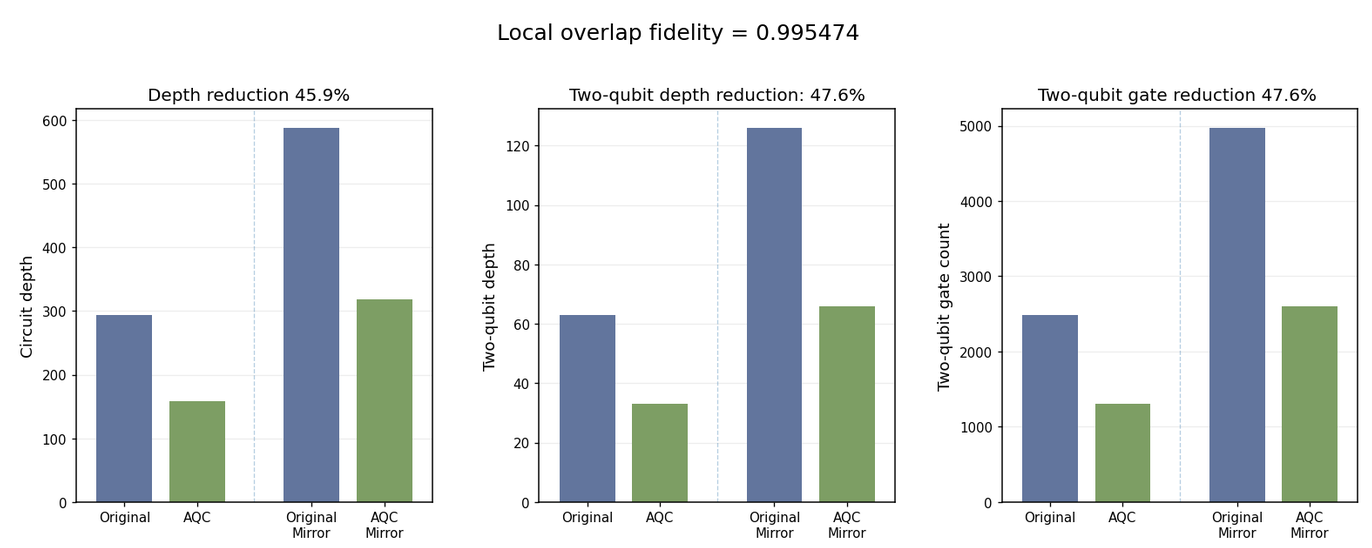}
\caption{Circuit-resource comparison for the original 10-repetition and AQCtensor-compressed 5-repetition circuits and their corresponding mirror circuits. Metrics are evaluated after transpilation using the same pass manager. AQCtensor reduces total circuit depth by 45.9\% and both two-qubit depth and two-qubit gate count by 47.6\%, while retaining a target-state fidelity of 0.995474.}
\label{fig:resource}
\end{figure}

Fig.~\ref{fig:resource} summarizes the circuit-resource reduction achieved by AQCtensor for the 80-qubit XYZ Heisenberg chain. The optimized target-state fidelity increased with the ansatz repetition count, from approximately 0 for $r=1$ and $2$ to 0.8594 for $r=3$, 0.9838 for $r=4$, and 0.995474 for $r=5$. Since $r=5$ was the shallowest ansatz to exceed the fidelity threshold of 0.99, it was selected as the compressed circuit, halving the repetition count from 10 to 5.

To quantify the hardware-level resource savings, the original and AQCtensor-compressed circuits were transpiled using the same physical-qubit layout and transpilation settings. The original circuit had a total depth of 294, a two-qubit depth of 63, and 2,487 two-qubit gates. After compression, these values were reduced to 159, 33, and 1,302, corresponding to reductions of 45.9\% in total depth and 47.6\% in both two-qubit depth and two-qubit gate count.
The corresponding self-mirrored circuits exhibited the same relative reduction in circuit resources, with total depth reduced by 45.9\% and both two-qubit depth and two-qubit gate count reduced by 47.6\%.

\subsection{Observable preservation on quantum hardware}
\label{sec:observables}

\begin{figure}[htbp]
\centering
\includegraphics[width=1.0\textwidth]{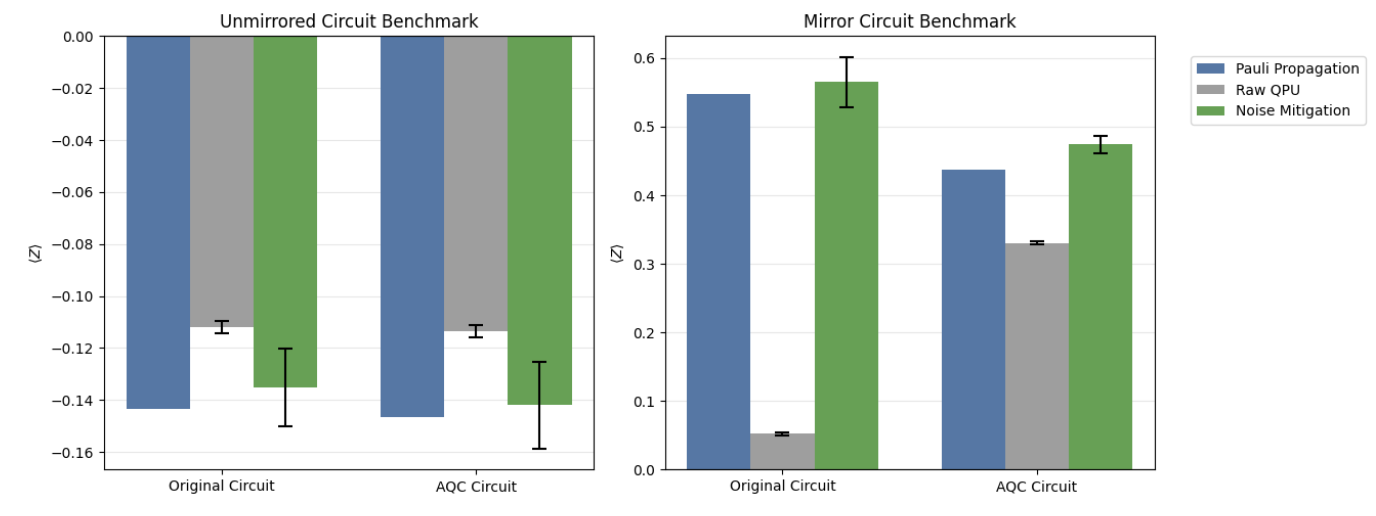}
\caption{Mean magnetization $\langle Z\rangle$ for the original and AQC-compressed circuits under unmirrored and mirrored benchmarks.}
\label{fig:z}
\end{figure}

We examined how AQCtensor compression affects observable estimates on quantum hardware. Fig.~\ref{fig:z} compares the mean magnetization $\langle Z \rangle$ obtained from raw and noise-mitigated QPU execution with the corresponding reference values for the original and compressed circuits, with the numerical values summarized in Table~\ref{tab:z}. For both circuits, noise mitigation shifts the measured $\langle Z \rangle$ values toward their respective references, and the two circuits show broadly comparable agreement after mitigation. In the corresponding mirror benchmark, the compressed circuit shows a smaller raw deviation from the corresponding Pauli-propagation estimate, indicating reduced accumulation of hardware errors.

\begin{figure}[htbp]
\centering
\includegraphics[width=1.0\textwidth]{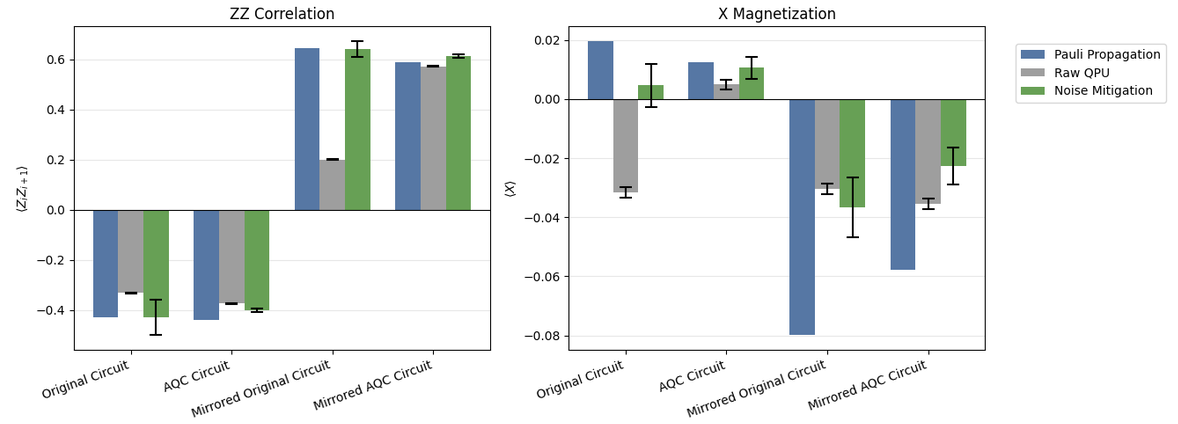}
\caption{Mean nearest-neighbor   correlation $C_{ZZ}$ (left) and mean $\langle X\rangle$ magnetization (right) for the original and AQCtensor-compressed circuits under unmirrored and mirrored benchmarks.}
\label{fig:zzx}
\end{figure}

Fig.~\ref{fig:zzx} extends the comparison to the mean nearest-neighbor correlation, $C_{\mathrm{ZZ}}$, and the mean transverse magnetization, $\langle X \rangle$, with the corresponding values summarized in Table~\ref{tab:zzx}. For $C_{\mathrm{ZZ}}$, the compressed circuit shows a smaller deviation from its reference in the raw QPU results, whereas the distinction for $\langle X \rangle$ is less pronounced because the corresponding reference values are small. After mitigation, both observables remain close to their respective references. The mirror measurements show the same qualitative reduction in raw hardware-error accumulation for $C_{\mathrm{ZZ}}$ and $\langle X\rangle$.

\begin{figure}[htbp]
\centering
\includegraphics[width=.60\textwidth]{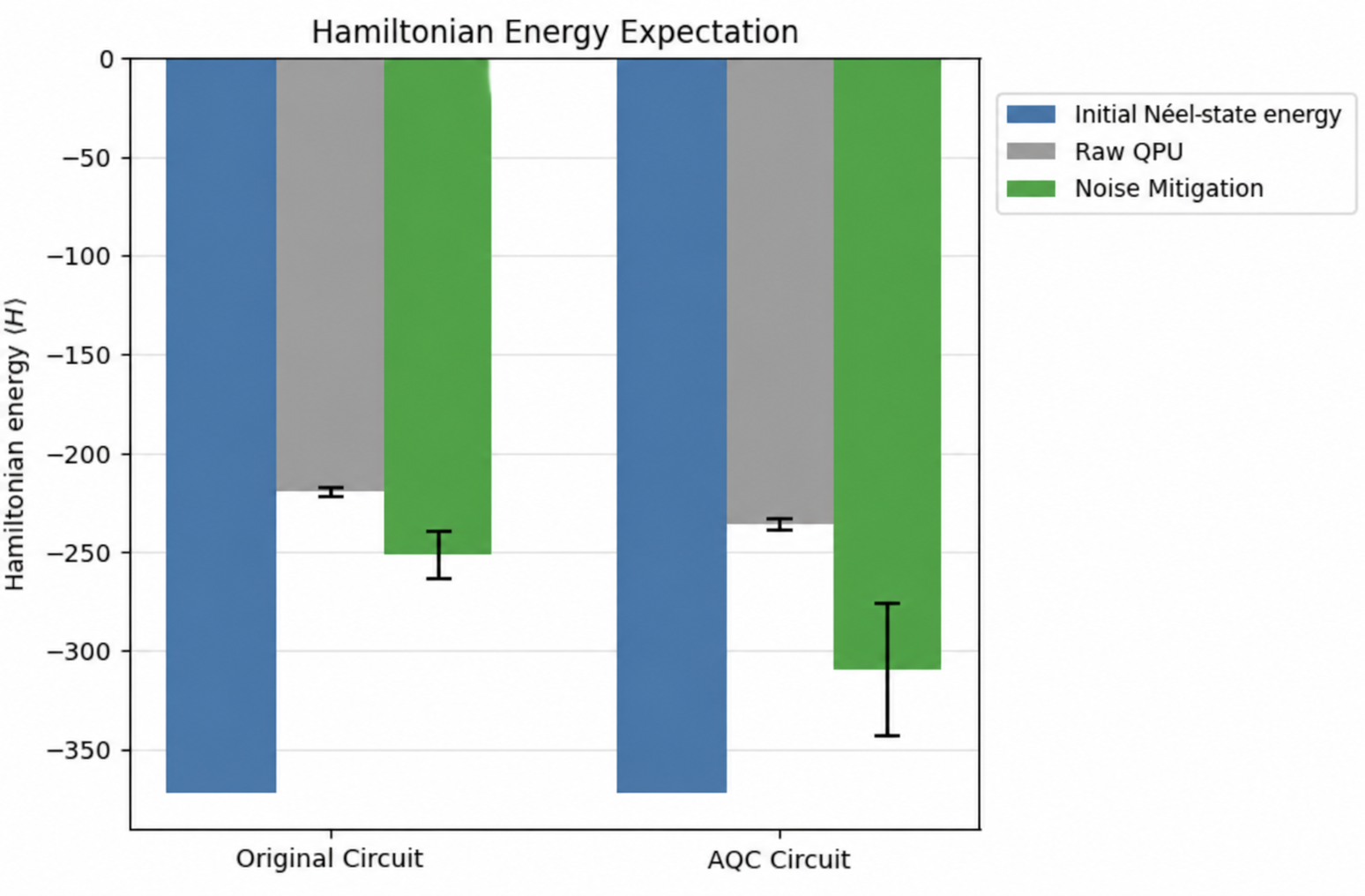}
\caption{Hamiltonian expectation values for the original and AQCtensor-compressed circuits, compared with the unevolved N\'{e}el-state reference.}
\label{fig:h}
\end{figure}

Fig.~\ref{fig:h} compares the Hamiltonian expectation values of the original and AQCtensor-compressed circuits with the initial N\'{e}el-state energy,
\[
E_0=\langle \phi_0 | H | \phi_0 \rangle .
\]
For the time-independent Hamiltonian considered here, the exact dynamics conserve the Hamiltonian expectation value,
\[
\langle \phi_0 | U^\dagger(t) H U(t) | \phi_0 \rangle = E_0 .
\]
The finite-step Trotter and AQC circuits introduce deviations from this conserved value, as reflected in their corresponding Pauli-propagated expectation values. Table~\ref{tab:h} summarizes these circuit-level values together with the raw and noise-mitigated QPU estimates.

The Pauli-propagated values remain close to $E_0$, whereas larger deviations are observed in the raw QPU estimates. Noise mitigation improves the agreement with the reference values for both circuits. Importantly, the AQCtensor-compressed circuit retains comparable agreement with the conserved energy while using considerably fewer circuit resources.

Overall, the observable-level results show that AQCtensor preserves the principal physical features of the original evolution across magnetization, spin correlation, and energy measurements, despite the substantial reduction in circuit resources.

\begin{table}[htbp]
\centering
\caption{Comparison of Pauli-propagation, QPU, and noise-mitigated $\langle Z\rangle$ magnetization estimates.}
\label{tab:z}
\scriptsize
\begin{tabular}{lllll}
\toprule
Observable & Circuit & Pauli Propagation & QPU & Noise Mitigation \\
\midrule
$\langle Z\rangle$ & Original & $-0.143350$ & $-0.111886 \pm 0.002468$ & $-0.135201 \pm 0.015038$ \\
$\langle Z\rangle$ & AQC Circuit & $-0.146375$ & $-0.113490 \pm 0.002487$ & $-0.141968 \pm 0.016805$ \\
$\langle Z\rangle$ & Mirror Original Circuit & $0.548247$ & $0.052112 \pm 0.002449$ & $0.564968 \pm 0.037131$ \\
$\langle Z\rangle$ & Mirror AQC Circuit & $0.437642$ & $0.330408 \pm 0.002254$ & $0.473912 \pm 0.012102$ \\
\bottomrule
\end{tabular}
\end{table}

\begin{table}[htbp]
\centering
\caption{Comparison of Pauli-propagation, QPU, and noise-mitigated  $C_{ZZ}$ and $\langle X\rangle$ estimates.}
\label{tab:zzx}
\scriptsize
\begin{tabular}{lllll}
\toprule
Observable & Circuit & Pauli Propagation & QPU & Noise Mitigation \\
\midrule
$\langle ZZ\rangle$ & Original & $-0.430104$ & $-0.331529 \pm 0.001707$ & $-0.429552 \pm 0.071930$ \\
$\langle ZZ\rangle$ & AQC Circuit & $-0.439385$ & $-0.373635 \pm 0.001697$ & $-0.401913 \pm 0.007782$ \\
$\langle ZZ\rangle$ & Mirror Original Circuit & $0.644866$ & $0.200425 \pm 0.001781$ & $0.642689 \pm 0.030792$ \\
$\langle ZZ\rangle$ & Mirror AQC Circuit & $0.590764$ & $0.572969 \pm 0.001505$ & $0.614403 \pm 0.007060$ \\
$\langle X\rangle$ & Original & $0.019573$ & $-0.031623 \pm 0.001782$ & $0.004594 \pm 0.007246$ \\
$\langle X\rangle$ & AQC Circuit & $0.012419$ & $0.004858 \pm 0.001787$ & $0.010595 \pm 0.003696$ \\
$\langle X\rangle$ & Mirror Original Circuit & $-0.079969$ & $-0.030483 \pm 0.001785$ & $-0.036747 \pm 0.010059$ \\
$\langle X\rangle$ & Mirror AQC Circuit & $-0.057827$ & $-0.035456 \pm 0.001787$ & $-0.022752 \pm 0.006268$ \\
\bottomrule
\end{tabular}
\end{table}

\begin{table}[t]
\centering
\caption{Comparison of the conserved-energy reference, Pauli-propagated circuit values, and raw and noise-mitigated QPU Hamiltonian expectation values.}
\small
\setlength{\tabcolsep}{4pt}
\begin{tabular}{lcccc}
\hline
Circuit &
$E_0$ &
Pauli Prop. &
QPU &
Mitigated QPU \\
\hline
Original &
$-372.278729$ &
$-370.387676$ &
$-219.484751 \pm 1.968351$ &
$-251.536910 \pm 12.177642$ \\
AQC Circuit &
$-372.278729$ &
$-369.579292$ &
$-235.554884 \pm 2.979196$ &
$-309.931387 \pm 33.402065$ \\
\hline
\end{tabular}
\label{tab:h}
\end{table}

\subsection{Process-level and local-state characterization}
\label{sec:preservation}

We next examined the relationship between the original and AQCtensor-compressed evolutions beyond individual observables using process-level and local-state diagnostics. These analyses were based on the relative circuit
\[
W = U_{\mathrm{original}}^\dagger U_{\mathrm{AQC}},
\]
which approaches the identity when the two ideal evolutions become equivalent.

\begin{figure}[htbp]
\centering
\includegraphics[width=.90\textwidth]{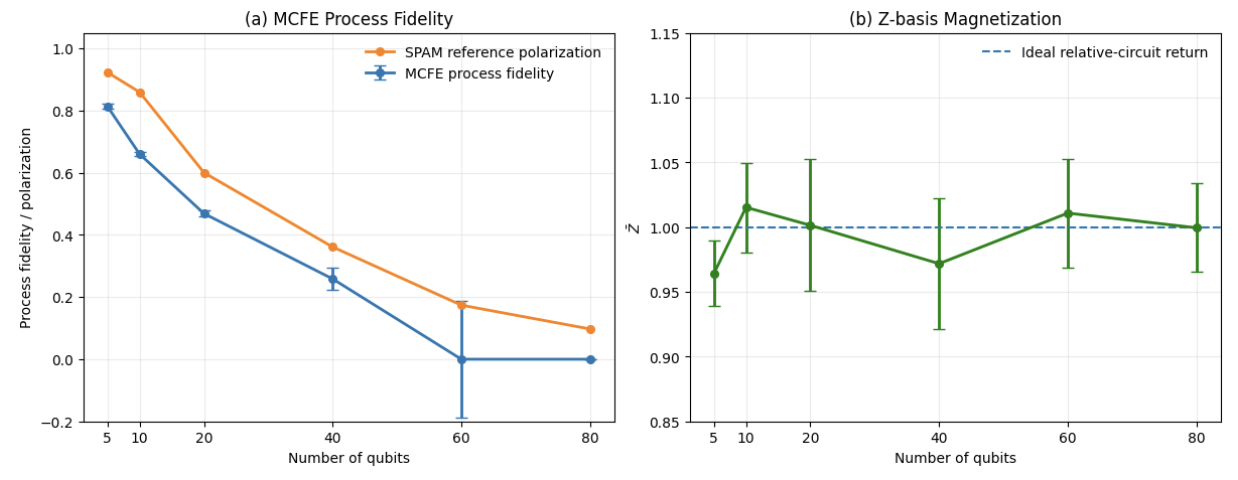}
\caption{Hardware-level comparison of the original and AQCtensor-compressed circuits using the relative circuit $W=U_{\mathrm{original}}^\dagger U_{\mathrm{AQC}}$. (a) MCFE process-fidelity estimatesand polarization as functions of system size. (b) Mitigated $N$-qubit-averaged $Z$-basis magnetization, with the dashed line indicating the ideal identity-return value $\langle Z\rangle = 1$.}
\label{fig:process}
\end{figure}

We first characterized the hardware implementation of the relative circuit using mirror-circuit fidelity estimation (MCFE) \cite{Proctor2021ScalableMirror,Erhard2019CycleBenchmarking,Proctor2022EstablishingTrust}. While full quantum process tomography is infeasible at this scale, MCFE provides a scalable means of estimating the process fidelity of the hardware implementation of $W$. As shown in Fig.~\ref{fig:process}(a), the MCFE estimates decrease from $0.813$ at 5 qubits to $0.258$ at 40 qubits, reaching the estimation boundary at 60--80 qubits. The SPAM-reference polarization, obtained from calibration circuits that exclude W, characterizes the effects of state preparation, measurement, and reference operations. It also decreases with system size, from 0.888 at 5 qubits to 0.043 at 80 qubits, but generally remains higher than the MCFE estimates for 10–80 qubits. These trends are consistent with increasing accumulated hardware errors at larger system sizes, where the calibration signal becomes weak.

To complement the MCFE analysis, we measured the $N$-qubit-averaged $\langle Z\rangle$ on the relative circuit as a basis-dependent return diagnostic. This quantity averages single-qubit expectation values and therefore does not quantify the probability of returning the entire system to the input state. Unlike the main compression experiment, which achieved a Néel-state fidelity of 0.995474, this experiment evaluates the relative circuit on the all-zero computational-basis input $|0\rangle^{\otimes N}$. For an ideal identity operation, the computational-basis input state is recovered and $\langle Z\rangle = 1$. Unlike MCFE, this quantity can be measured directly with the Qiskit Runtime Estimator and combined with error mitigation techniques. As shown in Fig.~\ref{fig:process}(b), the mitigated value remains close to 1 up to 80 qubits. The contrast with the lower MCFE estimates is consistent with accumulated hardware noise being a major contributor to the reduced execution fidelity, rather than necessarily indicating compression-induced errors.

We also evaluated the global return probability,
$P(0^{\otimes N})$, from computational-basis samples of the same relative-circuit construction using 4,000 Sampler shots. The return probability was already below 1\% at $N=10$, and no all-zero outcomes were observed for $N \geq 20$. This highlights the finite-shot sampling limitation of global return measurements at larger system sizes, in contrast to the Estimator-based expectation values that remain well resolved.

Finally, we compared the local-states produced by the original and
AQCtensor-compressed circuits using reduced-state tomography. Reduced density
matrices characterize the correlations and state structure of a subsystem
within a larger many-body state \cite{Xin2017RDM,Guo2023QuantumST}. For each
selected four-qubit region, the corresponding reduced density matrix was
reconstructed from local Pauli measurements.

\begin{figure}[htbp]
\centering
\includegraphics[width=1\textwidth]{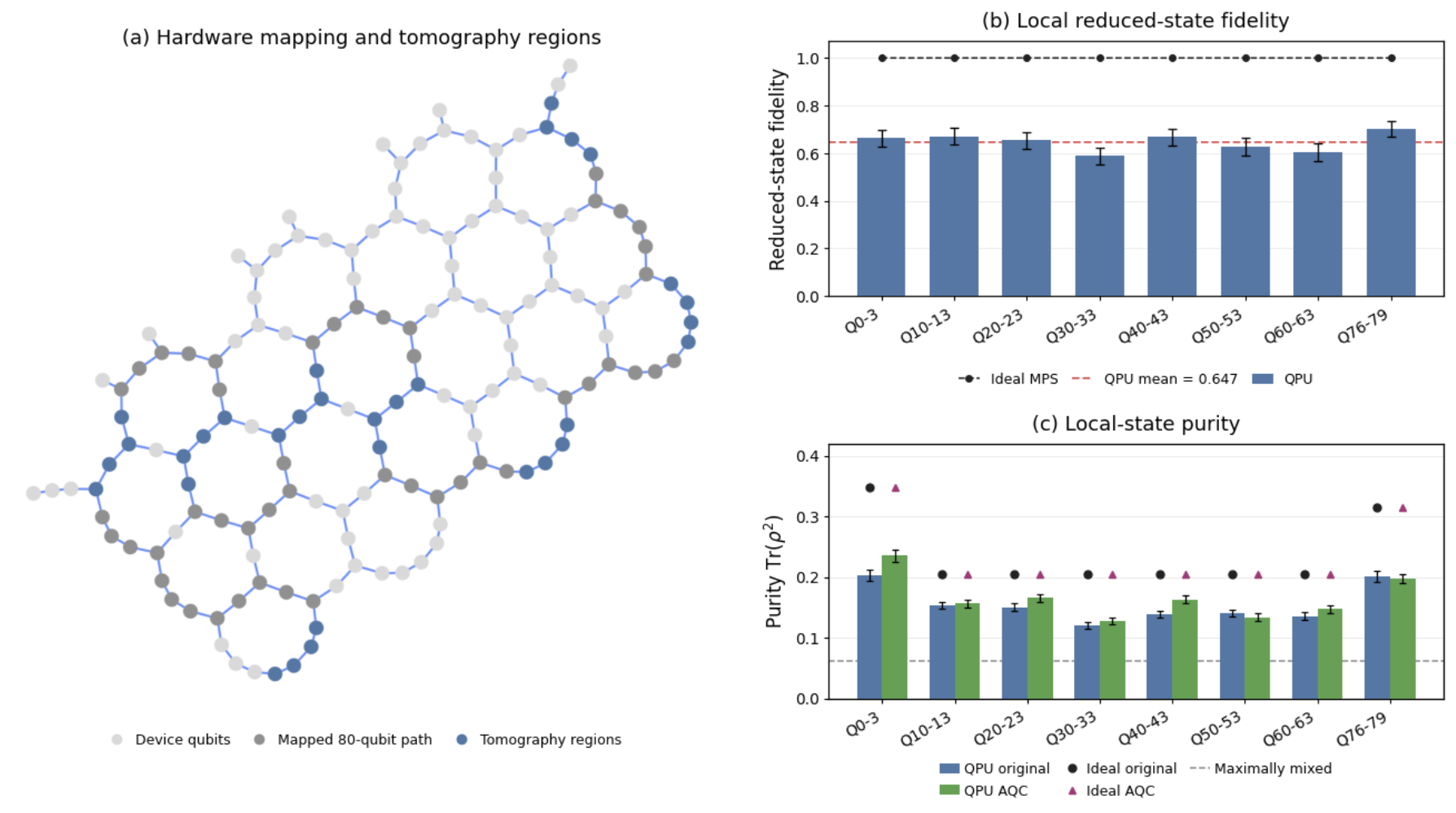}
\caption{
Local reduced-state analysis across the 80-qubit system.
(a) Physical mapping of the logical 80-qubit circuit on the Heron processor. Gray nodes indicate the mapped 80-qubit layout, and blue nodes indicate the eight four-qubit regions selected for reduced-state tomography.
(b) Reduced-state fidelity between the original and
AQCtensor-compressed circuits across the selected four-qubit
subsystems, comparing ideal MPS and hardware-reconstructed
states. The red dashed line indicates the mean QPU
reduced-state fidelity of 0.647.
(c) Local-state purity of the original and compressed
circuits, comparing ideal MPS references with the
hardware-reconstructed states across the same subsystems.
Purity is evaluated as $\mathrm{Tr}(\rho^2)$; the black
dashed line indicates the purity of a maximally mixed
four-qubit state, $1/16$.
}
\label{fig:local}
\end{figure}

Eight spatially separated four-qubit regions were examined across the mapped
80-qubit layout, including the boundary regions, $Q_{0-3}$ and $Q_{76-79}$,
together with representative interior regions, as shown in
Fig.~\ref{fig:local}(a). The tomography measurements used the same
error-mitigation protocol employed in the hardware analysis, including
dynamical decoupling, gate and measurement twirling, and linear zero-noise
extrapolation. As shown in Fig.~\ref{fig:local}(b), the reduced-state
fidelity between the hardware-executed original and AQCtensor-compressed
circuits ranges from approximately 0.59 to 0.70, with a mean of 0.647, and shows moderate
variation across the sampled regions. In comparison,
the corresponding ideal MPS reduced-state fidelities
remain close to unity.

The reconstructed local purities range from approximately
0.12 to 0.24 and remain broadly comparable between the
original and compressed circuits across the sampled regions,
as shown in Fig.~\ref{fig:local}(c). These values are lower
than the corresponding ideal MPS purities, reflecting the
increased mixedness of the hardware-reconstructed local
states. Together with the reduced-state fidelities in
Fig.~\ref{fig:local}(b), these results indicate moderate agreement between the hardware-reconstructed local states and broadly similar purities across the sampled regions, with lower fidelities and purities than the corresponding ideal MPS references.

\begin{table}[h]
\centering
\caption{MCFE execution-fidelity estimates and SPAM-reference polarization for the benchmark. Uncertainties are bootstrap standard deviations.}
\begin{tabular}{cccc}
\hline
Qubits & MCFE estimate & Std. & SPAM polarization \\
\hline
5  & 0.813354 & 0.008831 &                  0.922975 \\
10 & 0.659952 & 0.006378 &                  0.859244 \\
20 & 0.468697 & 0.010318 &                  0.600297 \\
40 & 0.258431 & 0.035827 &                   0.361136 \\
60 & 0.000000 & 0.186563 &                   0.173979 \\
80 & 0.000000 & 0.000000 &                    0.097271 \\
\hline
\end{tabular}
\end{table}

\section{Conclusion}
\label{sec:discussion}

Our results highlight two main findings. First, AQCtensor can reduce the hardware cost of large-scale quantum simulation while retaining physically relevant information. The compressed circuit requires substantially fewer quantum operations while maintaining, and in some cases improving, agreement with the reference values of the selected observables. Second, compressibility varies systematically with the underlying Hamiltonian. The nested-commutator measure is strongly associated with fixed-depth AQCtensor performance, while the directional response follows the symmetry of the anisotropic coupling structure. These trends are consistent with the relevance
of Hamiltonian noncommutativity emphasized by
commutator-scaling theories of Trotter error,
although the observed association with optimized
AQCtensor performance remains empirical.

Several limitations remain. The hardware study considers a single 80-qubit
XYZ Hamiltonian at one evolution time, and the local-state analysis samples
selected four-qubit regions rather than the complete 80-qubit state. MCFE
also becomes increasingly sensitive to accumulated hardware noise as the
diagnostic circuits grow, limiting process-level characterization at larger
system sizes. In addition, the error-mitigation settings were selected using
Pauli-propagated reference values and should therefore be interpreted as a
reference-guided comparison rather than an independent benchmark of
mitigation performance. The parameter sweep is also restricted to a limited
set of anisotropic couplings, field configurations, and product-formula
ans\"atze. The fixed-depth optimization was further performed with a finite
iteration budget, and symmetry-related coupling configurations did not always
yield numerically identical infidelities. These results should therefore be
interpreted in terms of broader trends across the Hamiltonian parameter space
rather than exact quantitative differences between symmetry-equivalent cases.

An important next step is to test whether the same compression trends persist in more challenging many-body settings. In particular, higher-dimensional systems such as two-dimensional spin Hamiltonians would provide a stringent test of whether the connection between Trotter-error structure and AQCtensor compressibility persists beyond one-dimensional chains. Another important direction is to determine whether these Trotter-error measures can be developed into practical indicators for identifying compressible regimes before performing the full AQC optimization. Such extensions would clarify both the generality and practical utility of the relationship observed here.

\section*{Acknowledgements}

 The authors acknowledge Ritajit Majumdar and Gwonhak Lee from the Quantum Algorithm Engineering team at IBM, and Somi Park from IBM Korea for fruitful discussions and contributions to achieving large-scale results on quantum hardware. This work was supported by the Korean ARPA-H Project through the Korea Health Industry Development Institute (KHIDI), funded by the Ministry of Health and Welfare (RS-2025-25456722); the Brain Pool Plus Fellowship Program, funded by the Ministry of Science and ICT (RS-2025-25427881); and the Artificial Intelligence Graduate School Program at Yonsei University (RS-2020-11201361).

\section*{Data Availability}
The data supporting the findings of this study are publicly available
on Zenodo at \url{https://zenodo.org/records/23786241}.

{\footnotesize
\bibliographystyle{unsrt}
\bibliography{references}
}

\clearpage

\section*{Supplementary Methods}

\subsection*{Supplementary Method A. Hamiltonian Configurations for Compressibility Analysis}

For the Hamiltonian-dependent compressibility analysis, we used the same open 80-qubit one-dimensional anisotropic XYZ Heisenberg chain and N\'{e}el initial state described in the main text. Each target evolution was generated using a 20-repetition second-order Suzuki--Trotter circuit at a total evolution time of $t=0.30$.

We considered two anisotropic coupling configurations,

\[
\mathbf{J}_1 = (0.5 \pi, 1.0 \pi, 1.5 \pi),
\quad
\mathbf{J}_2 = (1.0 \pi, 0.5 \pi, 1.5 \pi),
\]

which interchange $J_x$ and $J_y$ while keeping $J_z$ fixed. The external-field magnitude was varied over

\[
B \in \{0, 0.25, 0.50, 1.00, 2.00\}.
\]

For the single-axis field configurations, we used

\[
\mathbf{h}_x / \pi = (B,0,0),
\quad
\mathbf{h}_y / \pi = (0,B,0),
\quad
\mathbf{h}_z / \pi = (0,0,B),
\]

and for the equal-component two-axis configurations, we used

\[
\mathbf{h}_{xy} / \pi
=
(B/\sqrt{2}, B/\sqrt{2}, 0),
\]

\[
\mathbf{h}_{yz} / \pi
=
(0, B/\sqrt{2}, B/\sqrt{2}),
\]

\[
\mathbf{h}_{zx} / \pi
=
(B/\sqrt{2}, 0, B/\sqrt{2}).
\]

The factor $1/\sqrt{2}$ keeps the total field magnitude fixed when comparing single-axis and two-axis orientations.

For each Hamiltonian configuration, candidate AQCtensor ansätze with repetition counts from $r=1$ to $20$ were optimized against the corresponding target evolution. Compressibility was quantified in two ways. First, we recorded the minimum repetition count required to reach a target-state fidelity of at least $F \ge  0.99$. Second, we evaluated the optimized infidelity $1-F$ at the fixed shallow depth $r=4$. The first metric measures the circuit depth required to reproduce the target state, whereas the second provides a direct comparison of compression difficulty at the same ansatz depth. The MPS complexity was quantified by the maximum bond dimension of the target-state MPS, using a truncation cutoff of $10^{-13}$.

For the extended directional analysis in Fig. 2(c), we considered a broader family of anisotropic couplings beyond the two configurations used in Fig. 1. The normalized $x$- and $y$-couplings were parameterized as

\[
J_x/\pi = J_0 - \delta_J/2,
\quad
J_y/\pi = J_0 + \delta_J/2,
\]

with

\begin{align*}
J_0 &\in \{0.60, 0.75\},\\
J_z/\pi &\in \{0.75, 1.00, 1.25, 1.50\},\\
\delta_J &\in \{-0.8, -0.6, \ldots, 0.8\}.
\end{align*}

For each coupling configuration, we compared single-axis $x$- and $y$-directed fields at
$B \in \{1,2,3\}$, corresponding to
$\mathbf{h}_x/\pi=(B,0,0)$ and
$\mathbf{h}_y/\pi=(0,B,0)$.
The directional responses were quantified as

\[
\Delta C_2 = C_{2,y} - C_{2,x}
\]

and

\[
\Delta \chi_{\mathrm{MPS}} = \chi_y - \chi_x,
\]

where $\chi_x$ and $\chi_y$ denote the maximum target-state MPS bond dimensions for the $x$- and $y$-field evolutions, respectively.

\subsection*{Supplementary Method B. Detailed Procedures for Quantum Hardware Evaluation and Error Mitigation}

For the hardware-validation experiment, we used
$\mathbf{J}=(0.5 \pi,1.0 \pi,1.5 \pi)$ and
$\mathbf{h}=(0.5 \pi,1.0 \pi,2.0 \pi)$.
The target evolution was generated using a 10-repetition second-order Suzuki--Trotter circuit with a Trotter step of $\Delta t=0.03$, corresponding to a total evolution time of $t=0.30$.

This configuration was selected after evaluating the mean $\langle Z \rangle$ magnetization across several anisotropic-coupling and external-field settings using Qiskit pauli-prop \cite{pauli-prop}. Because the initial 80-qubit N\'{e}el state has zero mean magnetization, many tested configurations produced evolved values that also remained close to zero. We therefore selected a configuration with a more pronounced magnetization signal to improve the resolution of hardware-level differences between the original and compressed circuits. 

For hardware execution, the one-dimensional logical interaction graph was mapped onto a connected 80-qubit path using Qiskit's VF2Layout pass. The original and AQCtensor-compressed circuits were transpiled using the same physical-qubit layout and transpilation settings. The same mapping procedure was also used for their self-mirrored counterparts.

Pauli-propagation estimates for $\langle Z \rangle$, $\langle X \rangle$, and $\langle Z Z \rangle$ were obtained using the Qiskit pauli-prop package \cite{pauli-prop}. The measured observables were propagated backward through the circuit in the Heisenberg picture. Each circuit was decomposed into the supported gate set and separated into Clifford and non-Clifford components. Non-Clifford operations were propagated using the \texttt{propagate\_through\_circuit}, whereas Clifford operations were propagated exactly. After propagation to the initial computational-basis state, the expectation value was obtained from the coefficients of the Pauli terms containing only $I$ and $Z$ operators. 

To limit the classical cost of Pauli propagation, at most 500,000 Pauli terms were retained during propagation, and coefficients with magnitude below $10^{-12}$ were discarded. We verified the stability of the Pauli-propagation references across several retained-term budgets and observed only minor variations at large budgets. A maximum of 500,000 retained Pauli terms was therefore used for all reported reference values.

The reference for the Hamiltonian expectation value was evaluated separately. Rather than using Pauli propagation, we used the initial N\'{e}el-state energy

\[
E_0
=
\langle \phi_0 | H | \phi_0 \rangle.
\]

For the exact evolution generated by the time-independent Hamiltonian, this quantity is conserved. Because $|\phi_0 \rangle$ is a computational-basis state, the reference energy was evaluated directly from the known Hamiltonian by summing the expectation values of its Pauli terms.

QPU expectation values were evaluated using Qiskit Runtime EstimatorV2 with 4,000 shots per configuration. The error-mitigation pipeline combined twirled readout error extinction (TREX), $X_p X_m$ dynamical decoupling with ALAP scheduling, gate and measurement twirling, and probabilistic-error-amplification zero-noise extrapolation (PEA-ZNE) \cite{Cai2023QuantumEM}. 

For PEA-ZNE, we tested the three-point noise-factor sets

\[
(1.0,3.0,5.0),
\quad
(1.0,1.2,1.4),
\quad
(1.0,2.0,3.0),
\]

and the four-point sets

\[
(1.0,2.0,3.0,5.0),
\quad
(1.0,1.5,2.0,3.0),
\quad
(1.0,1.2,1.4,1.8).
\]

For each set of noise factors, linear, second-order polynomial, and exponential extrapolation models were evaluated. The mitigation configuration was selected separately for each circuit and observable by minimizing the absolute deviation from its corresponding reference value. For $\langle Z \rangle$, $\langle X \rangle$, and $\langle Z Z \rangle$, the Pauli-propagated expectation value was used as the reference, whereas $\langle H \rangle$ was compared with the exact initial-state energy defined above. The selected noise-factor sets and extrapolation models are summarized in Table~\ref{tab:mitigation-config}.

\begin{table}[htbp]
\centering
\caption{Selected noise-mitigation configurations used for the observable benchmarks.}
\label{tab:mitigation-config}
\scriptsize
\begin{tabular}{llll}
\toprule
Observable & Circuit & PEA noise factors & Extrapolator \\
\midrule
$\langle Z\rangle$ & Original & $(1.0,1.2,1.4)$ & Linear \\
$\langle Z\rangle$ & Compressed AQC & $(1.0,1.2,1.4)$ & Linear \\
$\langle Z\rangle$ & Original mirror & $(1.0,1.5,2.0,3.0)$ & Quadratic \\
$\langle Z\rangle$ & Compressed AQC mirror & $(1.0,2.0,3.0,5.0)$ & Linear \\
$\langle ZZ\rangle$ & Original & $(1.0,1.2,1.4,1.8)$ & Quadratic \\
$\langle ZZ\rangle$ & Compressed AQC & $(1.0,1.2,1.4,1.8)$ & Linear \\
$\langle ZZ\rangle$ & Original mirror & $(1.0,1.2,1.4)$ & Linear \\
$\langle ZZ\rangle$ & Compressed AQC mirror & $(1.0,2.0,3.0,5.0)$ & Linear \\
$\langle X\rangle$ & Original & $(1.0,1.2,1.4,1.8)$ & Linear \\
$\langle X\rangle$ & Compressed AQC & $(1.0,1.5,2.0,3.0)$ & Linear \\
$\langle X\rangle$ & Original mirror & $(1.0,2.0,3.0,5.0)$ & Quadratic \\
$\langle X\rangle$ & Compressed AQC mirror & $(1.0,1.5,2.0,3.0)$ & Linear \\
$\langle H\rangle$ & Original & $(1.0,1.2,1.4,1.8)$ & Linear \\
$\langle H\rangle$ & Compressed AQC & $(1.0,1.2,1.4)$ & Linear \\
\bottomrule
\end{tabular}
\end{table}

\subsection*{Supplementary Method C. Detailed Procedures for Process-level diagnostics and Local-State Analysis}

Process-level diagnostics were evaluated using the relative circuit

\[
W = U_{\mathrm{original}}^\dagger U_{\mathrm{AQC}},
\]

which approaches the identity when the original and compressed circuits implement the same evolution.

Mirror-circuit fidelity estimation (MCFE) was implemented using pyGSTi following mirror-circuit benchmarking procedures \cite{Nielsen2020pyGSTi,Proctor2021ScalableMirror,Proctor2022EstablishingTrust}. MCFE was evaluated for 5, 10, 20, 40, 60, and 80 qubits. For each system size, randomized bare-reference, reference-reference, and SPAM-calibration circuit families were generated using five mirror samples and five SPAM samples, with 4,000 shots per circuit. Hardware execution used dynamical decoupling together with gate and measurement twirling. Process fidelity was estimated from the calibrated survival probabilities of the three circuit families, and statistical uncertainties were obtained from 500 bootstrap resamples. The SPAM-reference polarization was calculated from
the reference-circuit measurement counts using
Hamming-distance-based effective polarization.

We also evaluated the $Z$-basis identity-return observable on prefixes of 5, 10, 20, 40, 60, and 80 qubits of an optimized 80-qubit relative circuit. Unlike MCFE, where AQC was independently optimized at each system size, these prefixes were obtained by retaining only gates acting entirely within the first $N$ logical qubits. This quantity is a Pauli expectation value and was evaluated using Qiskit Runtime EstimatorV2. Each measurement used 4,000 shots with measurement-error mitigation, linear ZNE at noise factors $(1.0,1.5,2.0)$, $X_p X_m$ dynamical decoupling, and gate and measurement twirling.

Local reduced-state tomography was performed on eight four-qubit regions distributed along the 80-qubit chain:

\[
Q_{0-3},
Q_{10-13},
Q_{20-23},
Q_{30-33},
Q_{40-43},
Q_{50-53},
Q_{60-63},
Q_{76-79}.
\]

For each selected region, the four-qubit reduced density matrix was reconstructed separately for the original and AQCtensor-compressed evolutions from local Pauli expectation values. The reconstruction used the complete local Pauli basis,

\[
\rho_A
=
\frac{1}{2^4}
\sum_{P\in\{I,X,Y,Z\}^{\otimes 4}}
\langle P \rangle P.
\]

with the identity contribution fixed to unity. Therefore, $4^4-1=255$ nonidentity Pauli expectation values were evaluated for each subsystem and circuit.

The tomography measurements comprised 16 Estimator
configurations, corresponding to eight regions and
two circuits, with 255 nonidentity Pauli observables
per configuration.
We used a default shot budget of 1,024,
measurement-error mitigation, linear zero-noise
extrapolation at noise factors $(1.0,1.5,2.0)$,
$X_pX_m$ dynamical decoupling, and gate and
measurement twirling.

The linearly reconstructed density matrices were
Hermitian-symmetrized and projected onto the
positive-semidefinite state space by setting negative
eigenvalues to zero and renormalizing the trace.
The total absolute weight of negative eigenvalues
before projection ranged from 0.193 to 0.391
across the reconstructed states.
Reduced-state fidelity was then evaluated between
the projected original and AQCtensor-compressed
states, together with their local purities,

\[
P=\mathrm{Tr}(\rho^2).
\]

For comparison with the ideal evolution, four-qubit
reduced density matrices were also obtained from the
MPS representations of the original and compressed
circuits by partial tracing over the remaining qubits.
The MPS simulations used a maximum bond dimension of
128 and a truncation cutoff of $10^{-10}$.

\subsection*{Supplementary Method D. Additional anisotropic coupling configuration}

To examine whether the compression trends observed in the main text are
specific to the coupling configurations with $J_z/\pi=1.5$, we performed an
additional AQCtensor analysis for a second anisotropic coupling pair,
\[
(J_x,J_y,J_z)/\pi=(1.5,1.0,0.5)
\]
and
\[
(J_x,J_y,J_z)/\pi=(1.0,1.5,0.5).
\]
The same system size, evolution time, field configurations, fidelity
threshold, and optimization protocol used in the main-text compressibility
analysis were retained.

\begin{figure}[htbp]
\centering
\includegraphics[width=.90\textwidth]{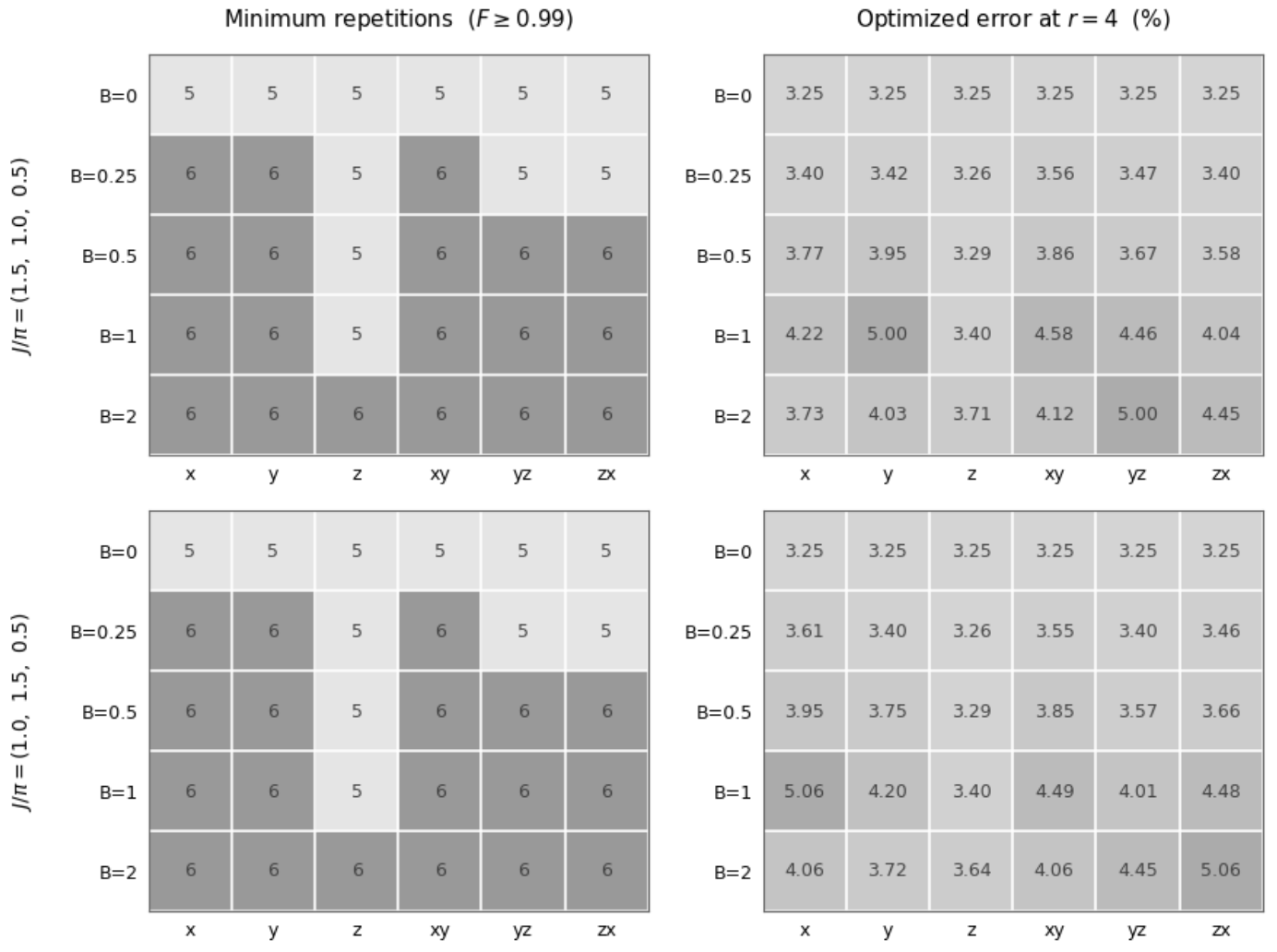}
\caption{
Additional AQCtensor compressibility analysis for anisotropic coupling
configurations with $J_z/\pi=0.5$.
The upper and lower rows correspond to
$(J_x,J_y,J_z)/\pi=(1.5,1.0,0.5)$ and $(1.0,1.5,0.5)$, respectively.
The left panels show the minimum repetitions for $F\geq0.99$, and the right panels show the infidelity at $r=4$, using the main-text optimization protocol.
}
\label{fig:supp_jz05}
\end{figure}

Figure~S\ref{fig:supp_jz05} summarizes the minimum number of repetitions
required to reach $F\geq0.99$ together with the achieved infidelity at fixed
depth $r=4$. Across the additional coupling pair, the minimum repetition count
is predominantly five or six, while the fixed-depth infidelity varies with
both field strength and field orientation. The directional variation is
generally weaker than that observed for the coupling set used in the main
text, indicating that the magnitude of the directional response itself
depends on the Hamiltonian parameters.

Exchanging $J_x$ and $J_y$ also changes the relative response to $x$- and
$y$-directed fields, although the finite-budget optimization does not yield
numerically identical values for all symmetry-related configurations.
Accordingly, these additional results provide a robustness check showing that compression
performance remains dependent on the combined Hamiltonian and field
configuration when $J_z$ is varied.

\subsection*{Supplementary Method E.
Comparison of Commutator Measures}
\label{supp:commutator_diagnostics}

To examine the relationship between Hamiltonian
noncommutativity and AQCtensor compressibility,
we use the even/odd/field nested-commutator measure
$C_2$ defined in the main text. This decomposition
reflects the Hamiltonian blocks of the implemented
Suzuki--Trotter formula and captures both bond--bond
and interaction--field noncommutativity.
We compare this measure with two alternatives to
distinguish the contributions of the Hamiltonian
splitting from those associated with the exact
leading-order Suzuki error.

First, the interaction--field-only measure is
defined as
\[
K=[H_{\mathrm{int}},H_{\mathrm{field}}],
\qquad
C_2^{\mathrm{IF}}=
\left(
\|[H_{\mathrm{int}},K]\|_{P,2}^{2}
+
\|[H_{\mathrm{field}},K]\|_{P,2}^{2}
\right)^{1/2},
\]
where $H_{\mathrm{int}}=H_{\mathrm{even}}+H_{\mathrm{odd}}$.
This measure captures interaction--field
noncommutativity but excludes bond--bond
contributions.

Second, we evaluate the leading
Baker--Campbell--Hausdorff (BCH) coefficient
of a reconstructed symmetric second-order
Suzuki--Trotter step $S_2(\Delta t)$,
which approximates the exact evolution
$e^{-iH\Delta t}$ over a single time step $\Delta t$:
\[
\log S_2(\Delta t)
=
-i\Delta t H+(\Delta t)^3L_3
+O((\Delta t)^5),
\qquad
C_{\mathrm{BCH}}=\|L_3\|_{P,2}.
\]
Here, $L_3$ denotes the leading-order error
coefficient arising from nested commutators
of the Hamiltonian terms, and
$C_{\mathrm{BCH}}$ measures its magnitude
using the Pauli-coefficient norm
$\|O\|_{P,2}=(\sum_P|c_P|^2)^{1/2}$
for $O=\sum_Pc_PP$.
Unlike $C_2$, the BCH coefficient accounts for
the ordering-dependent prefactors and cancellations
of the Suzuki--Trotter formula, thereby
characterizing the leading local error coefficient
rather than the aggregate commutator strength.

\begin{table}[htbp]
\centering
\caption{Pearson ($r$) and Spearman ($\rho$)
correlations between the three commutator
measures and fixed-depth AQCtensor infidelity
or saved maximum MPS bond dimension across
24 positive-field Hamiltonian configurations.}
\label{tab:supp_c2_comparison}
\small
\begin{tabular}{lcccc}
\hline
Measure & $r_{\mathrm{AQC}}$ & $\rho_{\mathrm{AQC}}$
& $r_{\mathrm{MPS}}$ & $\rho_{\mathrm{MPS}}$ \\
\hline
Even/odd/field $C_2$
& 0.887 & 0.900 & 0.744 & 0.454 \\
BCH coefficient $C_{\mathrm{BCH}}$
& 0.900 & 0.868 & 0.711 & 0.458 \\
Interaction--field $C_2^{\mathrm{IF}}$
& 0.948 & 0.950 & 0.761 & 0.468 \\
\hline
\end{tabular}
\end{table}

As shown in Table~\ref{tab:supp_c2_comparison},
all three measures exhibit positive correlations
with AQCtensor infidelity and MPS bond dimension.
The interaction--field measure yields the highest
pooled correlation with AQC infidelity, despite
omitting bond--bond contributions.

\begin{figure}[htbp]
\centering
\includegraphics[width=0.85\textwidth]{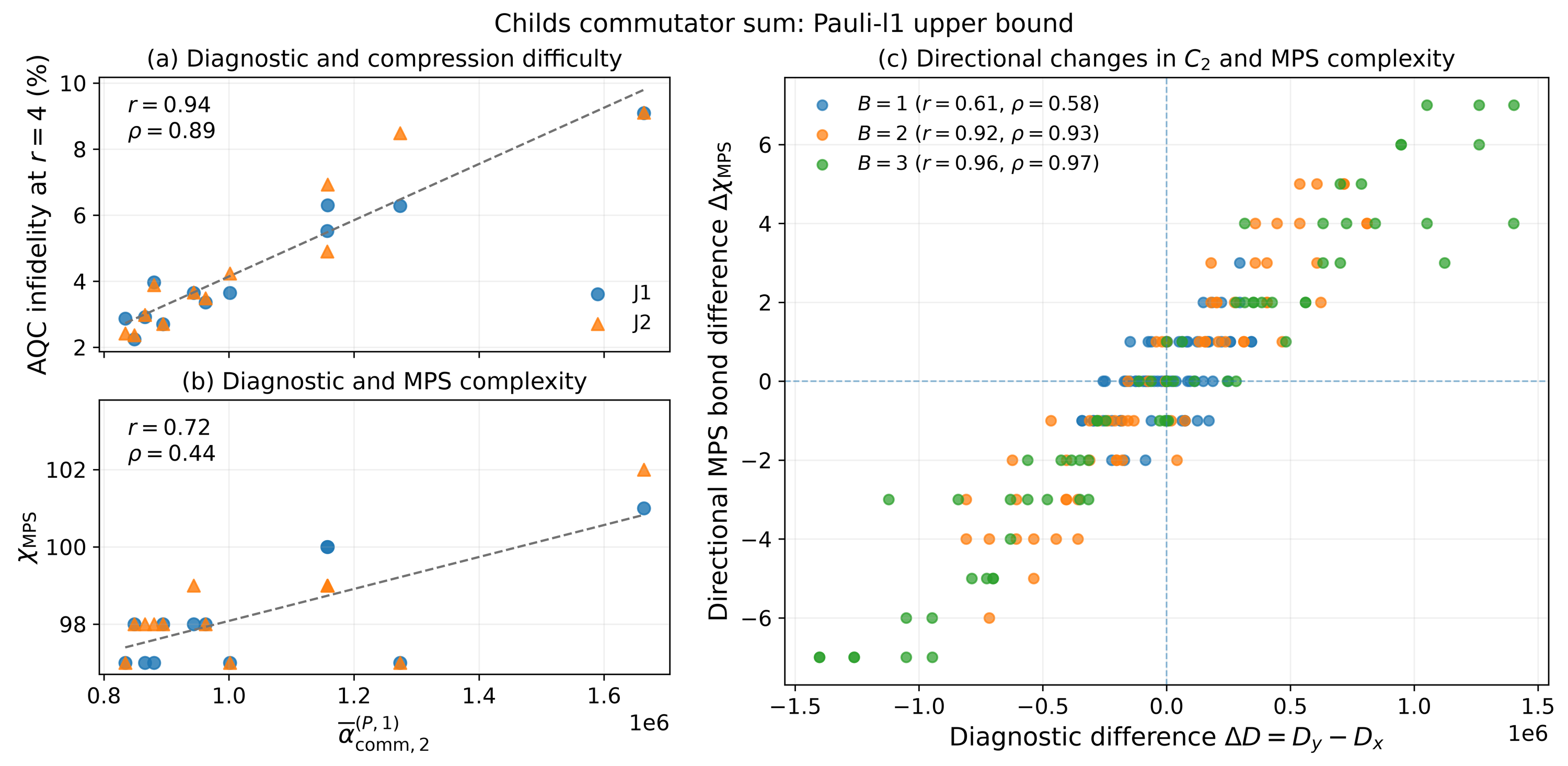}
\caption{
Comparison using the Pauli-$L_1$ upper bound
on the nested-commutator spectral-norm sum.
(a) Correlation with optimized AQCtensor
infidelity at $r=4$.
(b) Correlation with maximum MPS bond dimension.
(c) Directional differences in the commutator
measure and MPS bond dimension for
$x$- and $y$-directed fields at
$B=1$, $2$, and $3$.
}
\label{fig:supp_c2_childs}
\end{figure}

\begin{figure}[htbp]
\centering
\includegraphics[width=0.85\textwidth]{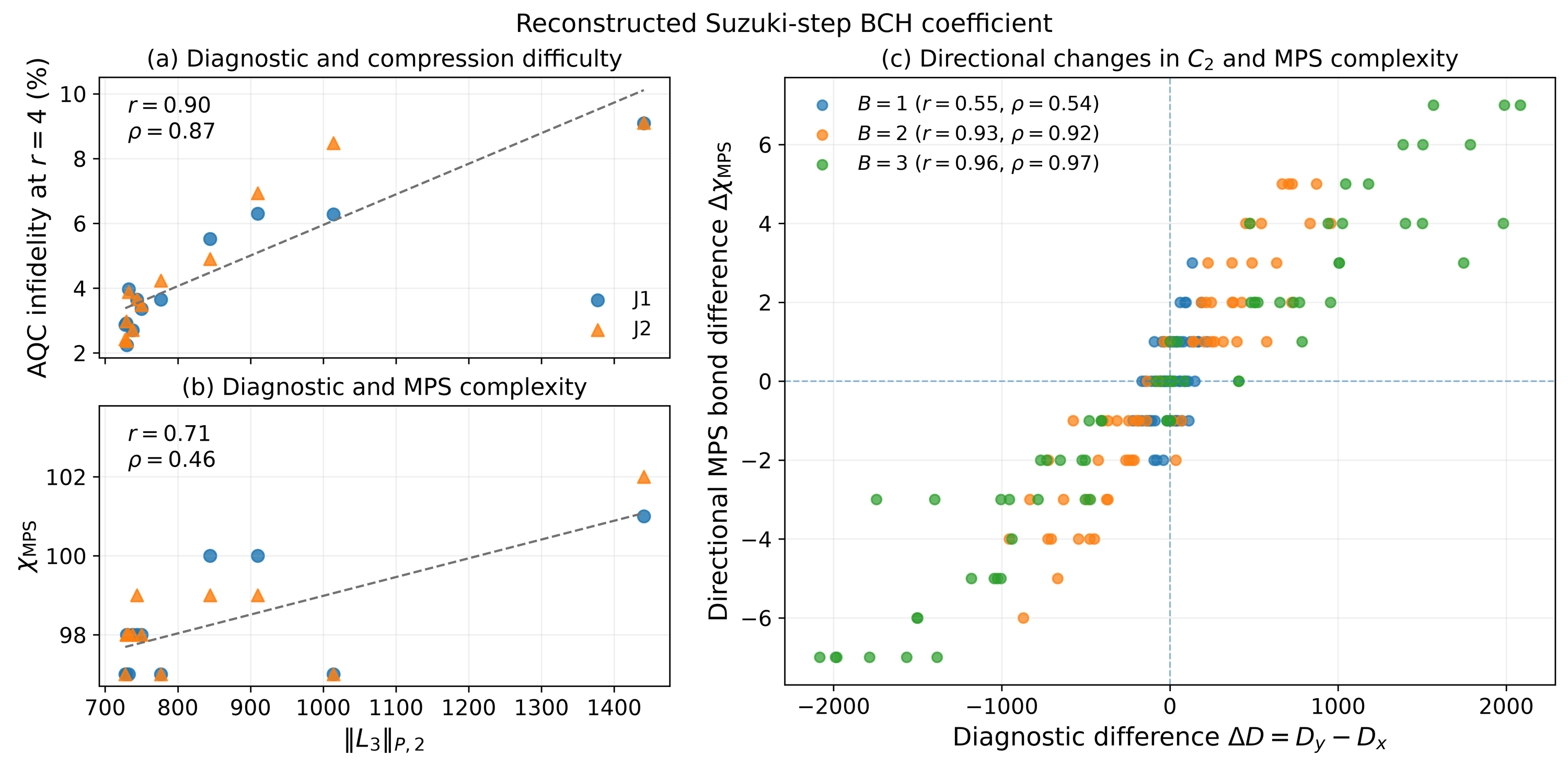}
\caption{
Comparison using the leading BCH coefficient
of a reconstructed second-order Suzuki step.
(a) Relation to optimized AQCtensor infidelity
at $r=4$.
(b) Relation to maximum MPS bond dimension.
(c) Correlation between the directional differences
in the BCH coefficient and MPS bond dimension
for $x$- and $y$-directed fields at $B=1$, $2$,
and $3$.
}
\label{fig:supp_c2_bch}
\end{figure}

\end{document}